# Giant Magneto-impedance in stress-annealed Finemet/Copper/Finemet based trilayer structures

B. Kaviraj[*] and F. Alves

LGEP/SPEE Labs, CNRS UMR 8507, Supèlec, Univ. Pierre et Marie Curie-P6, Univ. Paris Sud-P11, 11 rue Joliot-Curie, Plateau de Moulon, 91192 Gif-sur-Yvette, France

**Abstract**

The resistive and reactive components of magneto-impedance (MI) for Finemet/Copper/Finemet sandwiched structures based on stress-annealed nanocrystalline $Fe_{75}Si_{15}B_6Cu_1Nb_3$ ribbons as functions of different fields (longitudinal and perpendicular) and frequencies have been measured and analyzed. Maximum magneto-resistance and magneto-inductance ratios of 700% and 450% have been obtained in 30-600 kHz frequency range respectively. These large magneto-resistance and magneto-inductive ratios are a direct consequence of the large effective relative permeability due to the closed magnetic flux path in the trilayer structure. The influence of perpendicular bias fields ($H_{per}$) in the Longitudinal Magneto-impedance (LMI) configuration greatly improves the MI ratios and sensitivities. The maximum MI ratio for the resistive part increases to as large as 2500% for $H_{per} \sim 1$ Oe. The sensitivity of the magneto-resistance increases from 48%/Oe to 288%/Oe at 600 kHz frequency with the application of $H_{per} \sim 30$ Oe. Such high increase in MI ratios and sensitivities with perpendicular bias fields are due to the formation the favourable (transverse) domain structures.

[*] Corresponding author
Email : bhaskar.kaviraj@lgep.supelec.fr

## 1. Introduction

The giant magneto-impedance (GMI) effect [1, 2] in ultrasoft magnetic ribbons, films and wires is at the basis of advanced magnetic sensors with high sensitivity and high spatial resolution [3, 5]. The GMI effect is generally associated with a magnetic field-induced change in the complex impedance at high frequency [6, 7] and the change of the impedance is a direct consequence of the dependence of skin effect on relative magnetic permeability [8-11]. Interest in GMI was initiated in the early nineties when Panina et al. [2] and Beach et al. [12] reported a very large effect in amorphous ferromagnetic FeCoSiB wires at small magnetic fields and at relatively small frequencies. Since then, the GMI effect has been investigated in a variety of Fe- and Co-based amorphous ribbons [13-17], films [18-20], and wires [21-25]. GMI ratios ranging from a few % to several hundred % have been reported for amorphous Co-based wires with nearly zero magnetostriction. The permeability in soft amorphous magnetic materials depends on sample geometry, surface state [17, 26], orientation of the magnetic field, temperature, frequency, stress distribution in the material as well as the internal configuration of magnetization of the sample. Annealing in the presence of magnetic field or stress can prefer a preferential magnetization direction and release internal stresses which lead to much more important GMI effects [27-29].

Recently there has been active research on the GMI effect in sandwiched structures of F/M/F [30, 31] where F stands for a soft magnetic film and M for a non-magnetic layer with high conductivity. The GMI effect in these structures appears to be much stronger compared with that in the ferromagnetic single-layer films of same thicknesses. This is because the ac magnetic flux loop is closed and there is a weak influence of stray magnetic field in sandwiched structures. The condition of a strong skin effect for GMI effect is not required in a sandwiched structure as required in the case of single layered films. If the sandwich width is sufficiently large, the leakage of the magnetic flux induced by the current can be neglected [32]. Then the solution of the impedance equation for three-layered films infinitely extending in two directions can be used. Its low frequency expansion leads to the following: for $d_c \sigma_c >> d_m \sigma_m$, the inductive term proportional to $\mu_e$ can give the main contribution to Z, even in the case of a weak skin effect. Under this condition Z has a simple form:

$$Z = R_c \left[ 1 - 2j\mu_e \frac{d_m d_c}{\delta_c^2} \right] \qquad (1)$$

Here '$\sigma_c$' and '$\sigma_m$' represents the conductivities of the inner conductive and outer magnetic layers with thicknesses '$2d_c$' and '$d_m$' respectively and $\mu_e$ is the effective permeability of the magnetic layers in the transverse direction. The term $R_c = \dfrac{l}{2bd_c \sigma_c}$ represents the d.c resistance of the conductive lead, '$l$' being the length of the film where $b$ is the width of the magnetic layers. '$\delta_c$' represents the skin depth of the non-magnetic conductive layer expressed as $\delta_c = \sqrt{\dfrac{2}{\mu_0 \sigma_c \omega}}$ '$\omega$' being the angular frequency of the excitation current.

Expression (1) shows that the contribution of the magnetic layers to the sandwich impedance is described the external inductance with respect to the inner layer [32] and the external inductance has a linear dependence on $\mu_e$. In this case, an obvious impedance change can be obtained at frequencies lower than those required for a single ferromagnetic layer.

In this paper, we present the results of GMI effect in a stress-annealed FeNbCuSiB (F) based sandwiched structure which has been used as outer ferromagnetic layers and Cu is employed as an inner layer. It is well known that FeNbCuSiB has excellent soft magnetic properties such as high permeability (~ $10^5$ at 1 kHz in the ribbon), low coercive force and magnetostrictive constants [33]. Zhou *et al*. [34] studied the GMI effect in FeSiB/Cu/FeSiB trilayer films prepared by rf magnetron sputtering and found a maximum GMI ratio 17.2% for an applied field of 1600 A/m and at a frequency of 3 MHz. Xiao *et al*. [35] have investigated the GMI and domain structure in FeNbCuSiB single films and FeCuNbSiB/Cu/FeNbCuSiB trilayer structures of 6µm and 7µm thickness respectively. The magnetic films were investigated in the as-deposited and annealed states. A GMI ratio $\left( \frac{Z(H) - Z(H_{sat})}{Z(H_{sat})} \right)$ as large as 1733% was obtained in annealed trilayers at a frequency of 100 kHz. We have measured the GMI effect for the stress-annealed F/Cu/F sandwiched ribbons and obtained maximum magneto-resistance and magneto-inductance ratios of 700% and 450% respectively. The effect of a perpendicular dc bias field ($H_{per}$) has been found to increase the magnitudes of the MI ratios and sensitivities significantly. Maximum magneto-resistance as large as 2500% has been obtained with the application of $H_{per}$ ~1 Oe at 600 kHz frequency.

**2. Experiment**

Nanocrystalline Finemet ribbons of nominal composition $Fe_{75}Si_{15}B_6Cu_1Nb_3$ have been used as the magnetic layers. The ribbons were prepared by planar flow casting technique and purchased by Imphy Alloys. After short annealing under stress, $Fe_{75}Si_{15}B_6Cu_1Nb_3$ exhibit extraordinary linear and non-hysteretic characteristics with controlled transversal anisotropy field [36] and extreme low temperature dependence (0.17%/°C). Furthermore such annealed ribbons present the particularity to be ductile and are easily handled for the elaboration of F/Cu/F glued layered structure for GMI sensor. The layers of ribbons were 20 µm thick, 6 cm long and 1 cm wide. The thickness of the Cu strip was 40 µm. The applied dc field (H) and the perpendicular dc bias field ($H_{per}$) were provided by two pairs of Helmholtz coils. The applied dc field, H, was always along the direction of excitation current which was also along the longer dimension of the sandwiched films. This is the conventional longitudinal magneto-impedance (LMI) configuration. The direction of $H_{per}$ was transverse to that of H. The GMI measurements have been performed in the 30 kHz to 600 kHz frequency range with the help of a Lock-in-Amplifier (Model SR844, Stanford Research Systems). The relative changes in the resistive and inductive parts of GMI ratio have been expressed as $\delta R = \frac{R(H) - R(H_{sat})}{R(H_{sat})}$ and $\delta X = \frac{X(H) - X(H_{sat})}{X(H_{sat})}$ respectively where $H_{sat}$ is the saturating field where the impedance is reduced to very low values.

## 3. Results and Discussions

The field dependence of the relative change in the magneto-resistive (δR) and magneto-inductive (δX) components of impedance at different excitation frequencies have been depicted in Fig. 1. Very large magneto-impedance as functions of frequencies has been observed especially at low magnetic fields. It is worth pointing out here that single layered FeCuNbSiB films do not exhibit any measurable MI effects which underline the advantages of using multilayered films (reference [35], Xiao *et al*.). This is because the trilayer geometry allows a good magnetic flux closure leading to much higher effective permeability. At low frequencies, δR decreases monotonically with H and gets reduced to very low values at H = $H_{sat}$. At high frequencies (> 60 kHz), δR exhibits a peak and then decreases to very low values with the increase of H. The maximum value of this peak-field is around 15 Oe for all the frequencies. There is a dramatic increase in the maximum values of δR ($δR_{max}$) (from 191% at 30 kHz to almost 706%) with the increase in frequencies. The field dependence of δX as functions of frequency is different in two respects: first that it exhibit peaks even at lower frequencies and secondly there is a decrease in the maximum magneto-inductance ratio ($δX_{max}$) after f = 60 kHz where it decreases from 464% (60 kHz) to 195% (600 kHz).

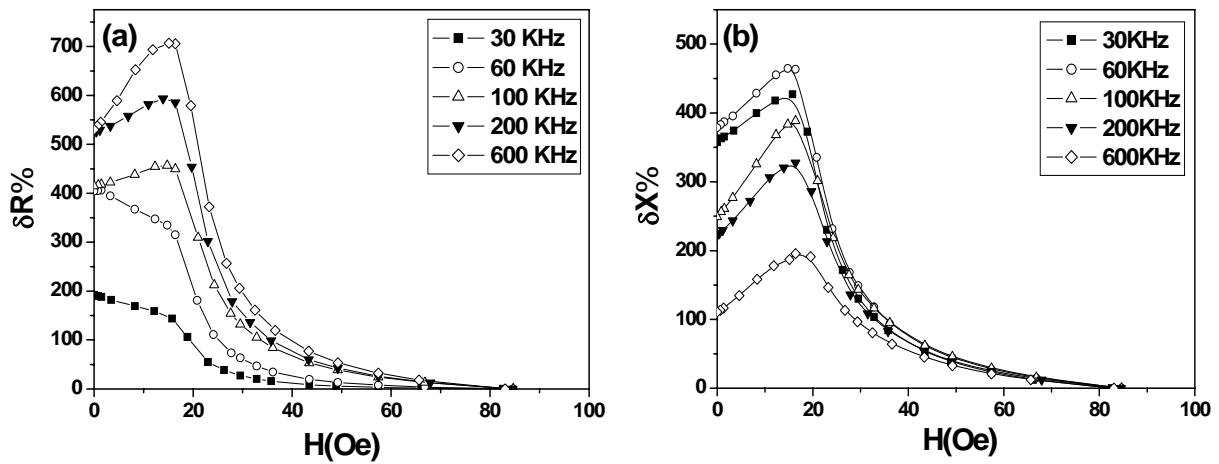

Fig. 1. Field dependence of magneto-resistance (a) and magneto-inductance (b) of the F/Cu/F trilayer structure at different excitation frequencies.

The frequency dependence of sensitivities (S) of R, X and Z where Z is the total impedance, has been depicted in Fig. 2. The sensitivity of the resistive component ($S_R$) is an increasing function of frequency. At low frequencies, $S_R$ increases almost linearly but it tends to saturate at higher frequencies (f > 600 kHz). The rise in sensitivity with the frequency is attributed to the increase in skin penetration depth, δ which increases the resistive component of Z. The sensitivity of the magneto-inductive component ($S_X$) rises to a maximum at f = 60 kHz and then decreases with the increase in frequency. The decrease in $S_x$ is due to the decrease of transverse permeability at higher frequencies. Maximum sensitivities of 48%/Oe and 30%/Oe have been obtained for R and X respectively.

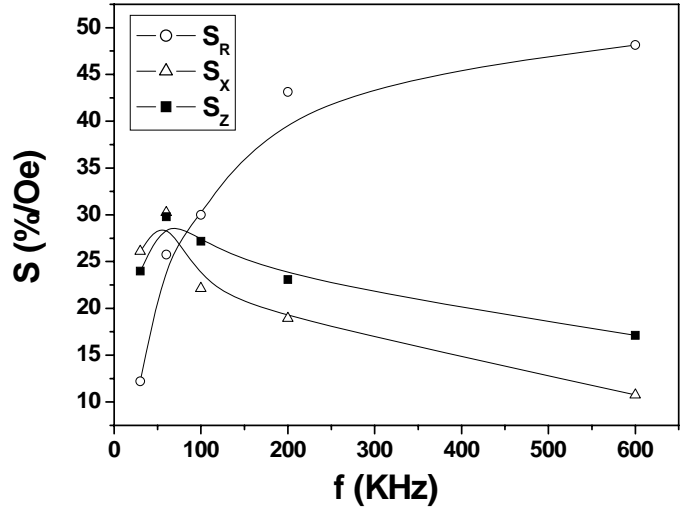

Fig. 2 Frequency dependence of sensitivities of R, X and Z of the F/Cu/F trilayer structure.

*3.1* **Effect of a perpendicular bias field**

In the LMI configuration, a bias field $H_{per}$ in a direction perpendicular to the external applied field H changes the equilibrium magnetization position and hence changes the impedance [6]. In our experimental condition, a perpendicular bias field $H_{per}$ was applied to notice any significant changes in the sensitivities or in the maximum MI ratios. The direction of $H_{per}$ was transverse to the direction of ac current and also to applied field H. The results of the field dependence of magneto-resistance and magneto-inductance at f = 600 kHz frequency have been depicted in Fig. 3.

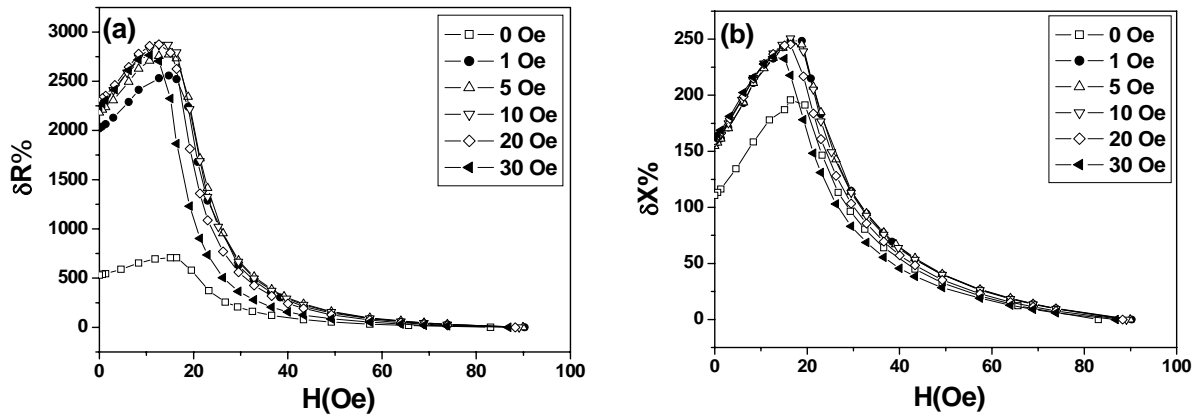

Fig. 3. Influence of perpendicular bias field $H_{per}$ on the resistive (a) and reactive (b) components of magneto-impedance.

The most striking observation is the exceptional high values of maximum MI ratios and sensitivities for $H_{per}$ ~1 Oe. In the absence of external field (H = 0), the maximum magneto-resistance ratio increases by almost four times to 2500% for $H_{per}$ just of the order of 1 Oe. The peaks in the resistive part become more prominent with the application of $H_{per}$. This confirms the presence of transverse anisotropy in the ribbons. The maximum magneto-resistance and magneto-inductance ratio increases to as much as 2876 % and 250% respectively for $H_{per}$ = 20 Oe. At bias fields $H_{per}$ > 20 Oe, the maximum MI ratios for both the resistive and inductive components start to decrease.

The sensitivity response of the resistive and inductive components of magneto-impedance as functions of different perpendicular bias fields and at 600 kHz frequency has been depicted in Fig. 4.

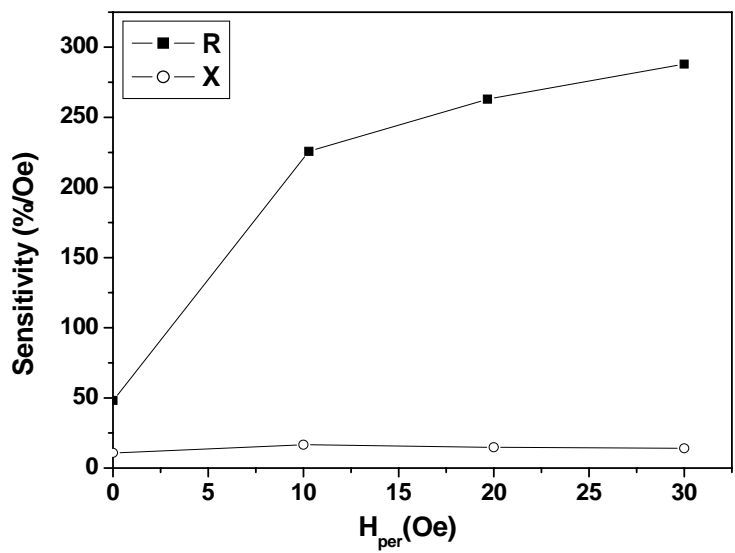

Fig. 4. Field sensitivities of the F/Cu/F multilayered structure under different perpendicular bias fields $H_{per}$ and at 600 kHz frequency.

Very high sensitivities especially in the resistive part have been obtained with the application of perpendicular bias fields up to 30 Oe where the sensitivity increased to almost 6 times when compared with sensitivity for $H_{per}$ = 0. The increase in sensitivity can be explained by the formation of a favorable domain structure along the transverse direction of the ribbon by the action of perpendicular bias fields. This increases the transverse permeability and hence the impedance of the ribbon. The direction of the external field, H being a hard direction causes the suppression of the transverse permeability thus decreasing the impedance causing the phenomenon of magneto-impedance. The higher susceptibilities, with the application of $H_{per}$ in the transverse direction, increase the maximum MI ratios with the application of external field H thus increasing the field sensitivities.

## 4. Conclusions

Magneto-impedance measurements have been carried out for stress-annealed F/Cu/F sandwiched structure based on nanocrystalline $Fe_{75}Si_{15}B_6Cu_1Nb_3$ ribbons in the frequency range of 30-600 kHz. Very high maximum MI ratios of about 700% and 450% have been observed in the resistive and reactive components respectively. This indicates that FeCuNbSiB films can be used as effective MI elements in a 'sandwiched' structure although no observable MI effects are found in single layer as-deposited films. This is because in a 'sandwiched' structure, the ac magnetic flux loop is closed and there is a weak influence of stray magnetic fields. While the sensitivity of magneto-resistance increased with frequency, the sensitivity of the magneto-inductance, on the other hand, showed an opposite behavior. Maximum sensitivities of 48%/Oe and 30%/Oe have been obtained for the magneto-resistance and magneto-inductance ratios respectively. A perpendicular bias field had a significant effect in improving the maximum MI ratios and sensitivities due to the formation of the favorable (transverse) domain structure. A very high sensitivity of 288%/Oe in the magneto-resistance ratio has been obtained with the application of $H_{per} \sim 30$ Oe at 600 kHz frequency. Such high sensitivities of MI especially in the presence of transverse fields enable the FeNbCuSiB sandwiched films a potential candidate in developing micro-magnetic sensors and magnetic heads for high density magnetic recording.

## References

[1] L.V. Panina, K. Mohri, K. Bushida, M. Noda, J. Appl. Phys. **76** (1994) 6198

[2] L.V. Panina, K. Mohri, Appl. Phys. Lett. **65** (1994) 1189

[3] T. Kitoh, K. Mohri, T. Uchiyama, IEEE Trans. Magn. **31** (1995) 3137

[4] K. Mohri, T. Uchiyama, L.V. Panina, Sens. Actuators A **59** (1997) 1

[5] M. Tewes, M. Lohndorf, A. Ludwig, E. Qandt, Tech. Mess. **68** (2001) 292

[6] L. Kraus, J. Magn. Magn. Mater. **195** (1999) 764

[7] D. Menard, Y. Yelon, J. Appl. Phys. **88** (2000) 379

[8] K.S. Byon, S.C. Yu, C.G. Kim, J. Appl. Phys. **89** (2001) 7218

[9] K.J. Jang, C.G. Kim, S.S. Yoon, S.C Yu, Mater. Sci. Eng., A **304** (2001) 1034

[10] W.S. Cho, H. Lee, S.W. Lee, C.O. Kim, IEEE Trans. Magn. **36** (2000) 3442

[11] J.F. Hu, L.S. Zhang, H.W. Qin, Y.Z. Wang, Z.X. Wang, S.X. Zhou, J. Phys. D **33** (2000) L45

[12] R.S. Beach, E. Berkowitz, Appl. Phys. Lett. **65** (1994) 3652

[13] S.U. Jen, Y.D. Chao, J. Non-Cryst. Solids, **207** (1996) 612

[14] H. Chiriac, F. Vinai, T.A. Ovari, C.S. Marinescu, F. Barariu, P. Tiberto, Mater. Sci. Eng., A **226** (1997) 646

[15] M. Vazquez, G.V. Kurlyandskaya, J.L. Munoz, A. Hernando, N.V. Dmitrieva, V.A. Lukshina, A.P. Potapov, J. Phys. I **8** (1998) 143

[16] F. Amalou, M.A.M. Gijs, J. Appl. Phys. **90** (2001) 3466

[17] E.E. Shalyguina, M. A. Komarova, V.V. Molokanov, C.O. Kim, C. Kim, Y. Rheem, J. Magn. Magn. Mater. **258** (2003) 174

[18] S.Q. Xiao, Y.H. Liu, L. Zhang, C. Chen, J.X. Lou, S.X. Zhou, G.D. Liu, J. Phys.: Condens. Matter **10** (1998) 3651

[19] S.Q. Xiao, Y.H. Liu, S.S. Yan, Y.Y. Dai, L. Zhang, L.M. Mei, Acta Phys. Sin. **48** (1999) S187

[20] J.Q. Wu, Y. Zhou, B.C. Cai, D. Xu, J. Magn. Mater. **213** (2000) 32

[21] M. Vazquez, M. Knobel, M.L. Sanchez, R. Valenzuela, A.P. Zhukov, Sens. Actuators A **59** (1997) 20

[22] D.X. Chen, J.L. Munoz, A. Hernando, M. Vazquez, Phys. Rev. B **57** (1998) 10699

[23] M.R. Britel, D. Menard, P. Ciureanu, A. Yelon, M. Rouabhi, R.W. Cochrane, C. Akyel, J. Gauthier, J. Appl. Phys. **85** (1999) 5456

[24] K. Mandal, S. Puerta, M. Vazquez, A. Hernando, Phys. Rev. B **62** (2000) 6598

[25] C. Gomez-Polo, M. Vazquez, M. Knobel, Appl. Phys. Lett. **78** (2001) 246

[26] F.L.A. Machado, A.E.P. de Araujo, A.A. Puca, A.R. Rodrigues, S.M. Rezende, Phys. Status, Solidi A **173** (1999) 135

[27] W.J. Ku, F.D. Ge, J. Zhu, J. Appl. Phys. **82** (1997) 5050

[28] G.V. Kurlyandskaya, J.M. Barandiaran, J. Gutierrez, D. Garcia, M. Vazquez, V.O. Vas'kovskiy, J. Appl. Phys. **85** (1999) 5050

[29] H.Z. Wu, Y.H. Liu, Y.Y. Dai, L. Zhang, S.Q. Xiao, Acta. Metall. Sin. **38** (2002) 1087

[30] K. Hika, L.V. Panina, K. Mohri, IEEE Trans. Magn. **32** (1996) 4594

[31] L.V. Panina, K. Mohri, T. Uchiyama, Physica A **241** (1997) 429

[32] A. Paton, J. Appl. Phys. **42** (1971) 5868

[33] Y. Yoshizawa, S. Oguma, K. Yamanchi, J. Appl. Phys. **64** (1998) 6044

[34] Y. Zhou, C.S. Yang, J.Q. Yu, X.L. Zhao, H.P. Mao, Chin. Phys. Lett. **17** (2000) 835

[35] S.Q. Xiao, Y.H. Liu, S.S. Yan, Y.Y. Dai, L. Zhang, L.M. Mei, Phys. Rev. B **61**, (2000) 5734; R.L. Sommer, C.L. Chien, Appl. Phys. Lett. **67,** (1995) 3346

[36] A-D. Bensalah, R. Barrué, F. Alves, L. AbiRached, Sensor Letters **5** (2007) 1